\documentclass[runningheads]{llncs}
\usepackage[T1]{fontenc}
\usepackage{graphicx}
\usepackage{subfigure}
\usepackage[inline]{enumitem}
\usepackage[nolist]{acronym}
\usepackage{xspace}
\usepackage[mode=text,detect-all,per-mode=symbol]{siunitx}
\usepackage{tikz}
\usepackage{booktabs}
\usepackage{amssymb}
\usepackage{color}

\newcommand{\red}[1]{{\color{red}{#1}}}

\newcommand{\secref}[1]{Sec.~\ref{#1}}
\newcommand{\tabref}[1]{Tab.~\ref{#1}}
\newcommand{\figref}[1]{Fig.~\ref{#1}}

\newcommand{\cf}{cf.\xspace}
\newcommand{\ie}{i.e.,\xspace}
\newcommand{\eg}{e.g.,\xspace}

\newcommand{\wrt}{w.r.t.\xspace}
\newcommand{\etal}{et~al.\xspace}

\newcommand*\circled[1]{\tikz[baseline=(char.base)]{
		\node[shape=circle,draw,inner sep=.5pt] (char) {#1};}}

\definecolor{diamond}{HTML}{9e2cf5}
\newcommand{\markdiamond}{{\color{diamond}{\scriptsize $\blacklozenge$}}\xspace}
\definecolor{star}{HTML}{ff7733}
\newcommand{\markstar}{{\color{star}{$\star$}}\xspace}

\begin{acronym}
	\acro{RF}{Random Forest}
	\acro{SVM}{Support Vector Machine}

	\acro{ICS}{Industrial Control System}
	\acro{IDS}{Intrusion Detection System}
	\acro{IIDS}{Industrial Intrusion Detection System}

	\acro{OCC}{One-Class Classifier}
\end{acronym}

\newcommand{ \evalSupervisedRFFinalPrecision }{ 0.998 } %
\newcommand{ \evalSupervisedRFFinalRecall }{ 0.987 } %
\newcommand{ \evalSupervisedSVMFinalPrecision }{ 0.885 } %
\newcommand{ \evalSupervisedSVMFinalRecall }{ 0.924 } %
\newcommand{ \evalMeasurementsSWaTSIMPLEMinMax }{ 10000 } %
\newcommand{ \evalMeasurementsSWaTInvariant }{ 703 } %
\newcommand{ \evalMeasurementsSWaTTABOR }{ 10000 } %
\newcommand{ \evalMeasurementsSWaTSeqSeqNN }{ 231 } %
\newcommand{ \evalMeasurementsWADISIMPLEMinMax }{ 10000 } %
\newcommand{ \evalMeasurementsWADIInvariant }{ 10000 } %
\newcommand{ \evalMeasurementsWADITABOR }{ 10000 } %
\newcommand{ \evalMeasurementsWADISeqSeqNN }{ 182 } %
\newcommand{ \evalMeasurementsBATADALSIMPLEMinMax }{ 10000 } %
\newcommand{ \evalMeasurementsBATADALInvariant }{ 1088 } %
\newcommand{ \evalMeasurementsBATADALTABOR }{ 10000 } %
\newcommand{ \evalMeasurementsBATADALSeqSeqNN }{ 500 } %
\newcommand{ \evalParametersSIMPLEMinMax }{ 2 } %
\newcommand{ \evalParametersInvariant }{ 10 } %
\newcommand{ \evalParametersTABOR }{ 7 } %
\newcommand{ \evalParametersSeqSeqNN }{ 6 } %
\newcommand{ \evalMaxSWaTPrecisionSIMPLEMinMax }{ 0.99 } %
\newcommand{ \evalMinSWaTPrecisionSIMPLEMinMax }{ 0.13 } %
\newcommand{ \evalMedianSWaTPrecisionSIMPLEMinMax }{ 0.89 } %
\newcommand{ \evalStdSWaTPrecisionSIMPLEMinMax }{ 0.23 } %
\newcommand{ \evalMedianSWaTFTABOR }{ 0.15 } %
\newcommand{ \evalMaxSWaTFTABOR }{ 0.79 } %
\newcommand{ \evalMedianSWaTFSIMPLEMinMax }{ 0.8 } %
\newcommand{ \evalMedianWADIFSIMPLEMinMax }{ 0.52 } %
\newcommand{ \evalMaxSWaTFSIMPLEMinMax }{ 0.82 } %
\newcommand{ \evalMaxWADIFSIMPLEMinMax }{ 0.55 } %
\newcommand{ \evalMedianSWaTRecallInvariant }{ 0.7 } %
\newcommand{ \evalMedianWADIRecallInvariant }{ 0.2 } %
\newcommand{ \evalMaxWADIFInvariant }{ 0.5 } %
\newcommand{ \evalTransferDiffSWaTWADIInvariant }{ -0.15 } %
\newcommand{ \evalTransferMedianSWaTWADIInvariant }{ 0.35 } %
\newcommand{ \evalMedianWADIFInvariant }{ 0.25 } %
\newcommand{ \evalTransferMedianSWaTWADITABOR }{ 0.23 } %
\newcommand{ \evalAvgDistUnsup }{ 0.18 } %

\begin{document}

\title{Deployment Challenges of Industrial Intrusion Detection Systems}
\author{
	Konrad Wolsing\inst{1,2} \and
	Eric Wagner\inst{1,2} \and
	Frederik Basels\inst{1} \and \\
	Patrick Wagner\inst{3} \and
	Klaus Wehrle\inst{2}
}
\authorrunning{K. Wolsing et al.}
\institute{
	Cyber Analysis \& Defense, Fraunhofer FKIE \email{\textit{first.last}@fkie.fraunhofer.de} \and
	COMSYS, RWTH Aachen University \email{\textit{last}@comsys.rwth-aachen.de} \and
	RWTH Aachen University \email{\textit{patrick.wagner3}@rwth-aachen.de}
}

\maketitle
\begin{abstract}
With the escalating threats posed by cyberattacks on Industrial Control Systems (ICSs), the development of customized Industrial Intrusion Detection Systems (IIDSs) received significant attention in research.
While existing literature proposes effective IIDS solutions evaluated in controlled environments, their deployment in real-world industrial settings poses several challenges.
This paper highlights two critical yet often overlooked aspects that significantly impact their practical deployment, i.e., the need for sufficient amounts of data to train the IIDS models and the challenges associated with finding suitable hyperparameters, especially for IIDSs training only on genuine ICS data.

Through empirical experiments conducted on multiple state-of-the-art IIDSs and diverse datasets, we establish the criticality of these issues in deploying IIDSs.
Our findings show the necessity of extensive malicious training data for supervised IIDSs, which can be impractical considering the complexity of recording and labeling attacks in actual industrial environments.
Furthermore, while other IIDSs circumvent the previous issue by requiring only benign training data, these can suffer from the difficulty of setting appropriate hyperparameters, which likewise can diminish their performance.
By shedding light on these challenges, we aim to enhance the understanding of the limitations and considerations necessary for deploying effective cybersecurity solutions in ICSs, which might be one reason why IIDSs see few deployments.
\end{abstract}

\section{Introduction}

The number of cyberattacks on \acp{ICS}, ranging from manufacturing over power grids to water and gas distribution, exploded in recent years~\cite{Alladietal2020Industrial,humayed2017cyber}.
The protection of such facilities is, however, not trivial as many systems rely on insecure legacy communication protocols, replacement of which is cumbersome, expensive, and often unrealistic due to high uptime requirements~\cite{Galloway2012introduction}.
Consequently, recent research focuses on easily retrofittable \acp{IIDS} specifically designed to take advantage of the unique characteristics of each \ac{ICS} by searching for anomalous behavior in largely predictable networking patterns and physical processes~\cite{Lamberts2023SoK}.

The foundation of these detection mechanisms is mostly rooted in classical \emph{supervised} machine-learning or \emph{\acp{OCC}}~\cite{Lamberts2023SoK,Wolsing2022IPAL}.
In supervised approaches, the \ac{IIDS} is trained on labeled samples of genuine behavior \emph{and} attacks to learn classifiers, \eg \acp{RF} or \acp{SVM}~\cite{Perezetal2018Machine}.
Meanwhile, \acp{OCC} are trained only on genuine \ac{ICS} behavior, \eg to identify the operational boundaries of physical measurements~\cite{Wolsing2022Can}, and deviations from this learned behavior are classified as potential attacks.

Research demonstrates the alleged effectiveness of hundreds of newly proposed \acp{IIDS} by evaluating them on dedicated datasets and publishing achieved detection performances~\cite{Contietal2021A,Lamberts2023SoK}.
In vitro, these \acp{IIDS} achieve excellent results~\cite{Feng2019A,Kim2020Anomaly,Lin2018TABOR,Wolsing2022Can}.
However, when it comes to real-world deployments, these solutions are challenging to configure~\cite{Erbaetal2020No} and then cannot perform as promised~\cite{Ahmedetal2020Challenges,Sommeretal2010Outside}.
Consequently, the performance derived by current evaluation methodologies seems hardly representative of the actual quality of an \ac{IIDS} if deployed in the real world.
While the scientific literature already identifies various challenges for transferring \acp{IIDS} from research into practice~\cite{Ahmedetal2020Challenges,Sommeretal2010Outside}, we proclaim that two crucial aspects impacting \acp{IIDS}' deployability remain unaddressed.

First, it remains unclear how much training data is required to maximize detection performance.
This question is especially critical in the case of supervised \acp{IIDS}, where the collection of attack samples in a testbed might still be relatively easy, but collecting real-world attack samples is much harder~\cite{Baderetal2023Comprehensively}.
\acp{OCC}' genuine training data, on the other hand, is easily collectable, but they still require hyperparameter tuning~\cite{hutter2019automated}.
Yet, hyperparameter tuning is rarely intuitive, especially with often-employed custom classifiers, and it remains unknown whether it is possible to transfer good hyperparameters between \ac{ICS} deployments as considered feasible in other machine-leraning domains~\cite{probst2019tunability}.
In research, the authors thus may optimize them for a given dataset (with attacks), which is, however, unfeasible in practice due to lack of attack samples.

Machine-learning for intrusion detection in the \ac{ICS} domain is especially challenging because of hard to obtain attack samples from cyberattacks in real systems, as their collection would expose, disrupt, and potentially damage sensitive critical infrastructure and manufacturing facilities.
For artificial datasets and testbeds as used in research~\cite{Contietal2021A}, on the other hand, it is relatively easy to generate such attack samples.
Thus far, \ac{IIDS} proposals do, however, all require custom training phases for the concrete deployment with hardly any model transferability across scenarios~\cite{Etalle2017From,Wolsing2022IPAL}.
We thus observe a large discrepancy between training data availability for research activities and real-world deployments, which may be the culprit for the reported challenging deployment of current research proposals~\cite{Ahmedetal2020Challenges}.

\textbf{Contributions.} To investigate the potential influence of training data availability on \acp{IIDS}' deployability, we make the following contributions:
\begin{itemize}[label=\textbullet,leftmargin=3mm, topsep=5pt]
	\item We demonstrate that the amount of attack samples in training significantly influences the performance of \acp{IIDS} based on supervised machine-learning.

	\item We show that the influence of hyperparameters for \ac{OCC}-based \acp{IIDS} varies tremendously.
	While some may depend on attack samples for tuning, others are largely hyperparameter-agnostic and even generalize across deployments.

	\item Based on our findings, we advocate for more expressive \ac{IIDS} evaluation procedures to close the gap between research and real-world \ac{IIDS} deployments.
\end{itemize}

\textbf{Availability Statement.}
To facilitate further research, we publish our data, scripts, and configurations to replicate our experiments from this publication: \url{https://zenodo.org/records/10728074}

\section{Background on Industrial Intrusion Detection}
\label{sec:bg:iids}

For readers unfamiliar with the topic of \emph{industrial} intrusion detection, we motivate the rationale of retrofitting detective solutions to \ac{ICS} and present one \ac{IIDS} from the literature and evaluated in this publication in the following in detail.

\acfp{ICS} resemble the foundation of many modern applications ranging from manufacturing, over the production and distribution of water, gas, or electricity, to autonomous vehicles~\cite{humayed2017cyber}.
Besides this diversity, one typical architecture that all these applications rely on are digital control loops measuring the environment with sensors and influencing it through actuators usually interconnected with industrial control networks~\cite{Galloway2012introduction}.
Consequently, \acp{ICS} are likewise susceptible to regularly occurring threats from cyberspace~\cite{Alladietal2020Industrial}, which can ultimately cause harm to the physical processes, businesses, and environment.
For their mitigation, either preventive measures such as authenticated and encrypted communication channels~\cite{Dahlmannsetal2022Missed} or detective approaches like \acp{IDS}~\cite{Lamberts2023SoK} can be implemented.
This publication focuses on the latter, which aim at timely indicating malicious behavior to \ac{ICS} operators before actual harm can be conducted and avoid attacks remaining uncovered.

To detect unwanted behavior, the detection methodologies underlying \emph{industrial} \acp{IDS} make great use of domain knowledge and \ac{ICS}-specific behavior~\cite{Wolsing2022IPAL}.
One key attribute is \acp{ICS}' notorious predictability, as they usually perform repetitive tasks~\cite{Galloway2012introduction}.
Based on a set of training data, note that supervised \acp{IIDS} require attack samples while \ac{OCC} methods solely train on benign data, a detection model can be trained and tuned with hyperparameters to indicate unexpected deviations, such as cyberattacks.
The goal of each approach and their tuning is to detect as many cyberattacks as possible while emitting few false positive alerts, which would have to be falsified by operators afterward.
The performance of an \ac{IIDS} is ultimately measured with metrics~\cite{Lamberts2023SoK} like the F1 score.

\begin{figure}[t]
	\centering
	\includegraphics[width=.8\columnwidth]{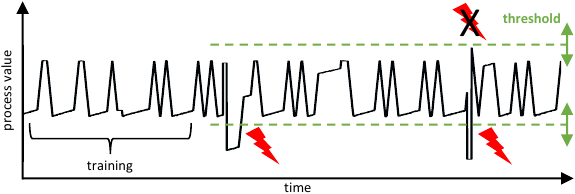}

	\caption{
	An \ac{IIDS} learns the repetitive patterns of an \ac{ICS}'s behavior to indicate anomalies.
	This requires finding a suitable hyperparameter, such as the threshold for MinMax visualized here~\cite{Wolsing2022Can}, which influences the alert decision of an \ac{IIDS}.}
	\label{fig:minmax-example}
\end{figure}

One approach to implementing such an  \ac{OCC}-based \ac{IIDS}, which serves as our introductory example, is MinMax (\cf \figref{fig:minmax-example})~\cite{Wolsing2022Can}.
The detection methodology is based on the fact that physical values measured by sensors usually reside within precise limits, \eg a boiler inside an \ac{ICS} has a lower and upper operational temperature.
MinMax first extracts these limits from a set of benign training data.
Then, since physical measurements can underlie natural variation and noise, the approach enlarges these limits by a configurable hyperparameter to avoid too many false positive alarms.
In the end, an alarm is raised if a measurement exceeds or undercuts the trained threshold.
Finally, the \ac{ICS} operators are in charge of analyzing the raised alarm and initiating countermeasures.

For \ac{OCC}-based \acp{IIDS}, as depicted here, the training requires benign data recorded, \eg during normal \ac{ICS} behavior.
Still, for deployment, hyperparameters, \ie the threshold, have to be adequately selected to reduce the number of false-positives and not miss attacks (\cf \figref{fig:minmax-example}).
Contrary, while supervised \acp{IIDS} can find adequate hyperparameters themselves during training as they also learn on malicious samples, obtaining this malicious data in \ac{ICS} is much more challenging as it involves actual physical processes~\cite{Baderetal2023Comprehensively}.

\section{Open IIDS Deployment Challenges in ICS}

After a short primer on \acp{IIDS}, we now highlight deployment challenges of \acp{IIDS} along recent related work, reproducibility studies, and meta-reviews (\secref{sec:problem:rw}).
Afterward, we formulate the research questions addressed in this paper (\secref{sec:problem:rq}).

\subsection{Related Work}
\label{sec:problem:rw}

For \ac{IDS} research, there exists a body of meta-studies that critically reflect the effectiveness and suitability of research proposals.
In that regard, Sommer \etal~\cite{Sommeretal2010Outside} argue, not specifically focusing on industrial networks, that machine-learning is better suited for finding similarities than differences, which complicates their application in anomaly detection.
Moreover, it is challenging to conduct sound evaluations, which they presume to be the reason why most approaches cannot keep up with expectations in real deployments. 
Adding to these issues, Ahmed \etal~\cite{Ahmedetal2020Challenges} identify scalability, exhaustive system modeling during training, and noisy input data as challenges seldom evaluated in live deployments.
Moreover, operational drift and component aging that change normal behavior become only apparent in real deployments~\cite{M.-R.etal2021Machine}.
However, the differences between training of (industrial) IDSs in artificial scenarios and real deployments have thus far not been analyzed.

Moreover, general machine learning research has examined the importance of hyperparameter tuning~\cite{hutter2019automated}.
Here, we are mostly concerned with second-level hyperparameters, \ie hyperparameters that must be set prior to training~\cite{probst2019tunability}.
To obtain a general understanding of the tunability of these second-level hyperparameters, Probst~\etal~\cite{probst2019tunability} analyzed six supervised machine-learning algorithms.
They found good default values working on many datasets and identified those hyperparameters worth considering for tuning.
In a similar study, Weerts~\etal~\cite{weerts2020importance} found out that leveraging default hyperparameters was non-inferior to tuning them.
However, all these works mostly consider supervised machine-learning and neither look at \ac{OCC} nor tackle the peculiarities in \ac{ICS}.
Regarding the latter, default values found in these works do not apply to the entirely different and custom \ac{OCC}-based \ac{IIDS} algorithms usually found in \ac{ICS} research.
Focusing on \ac{ICS}, Fung~\etal~\cite{Fungetal2022Perspectives} show exemplarily that three considered \acp{IIDS} deliver mostly stable performance under different hyperparameters.
However, the set of tested hyperparameters is derived from attack samples, which may not be available (in high quantity or quality) for real deployments. 

\subsection{Research Questions}
\label{sec:problem:rq}
The deployment of \acp{IIDS} in real industrial networks proves challenging, with experimental deployments failing to keep up with promising results from artificial scenarios.
We suspect training data availability, especially samples of attacks, to be one potential culprit for this situation.
Detection algorithms themselves are often applicable to multiple industrial domains~\cite{Wolsing2022IPAL}.
However, they assume to be trained separately for each deployment to learn the expected behavior.
For example, the learned boundaries of a water tank's maximum acceptable fill level differ for each \ac{IIDS} deployment.
Consequently, it is inevitable to train an \ac{IIDS} for a specific target use case.
Yet, this challenge of training an \ac{IIDS} is not critically reflected in research where simply another (existing) dataset can be leveraged.
To verify our suspicion and improve future evaluation methodologies of \acp{IIDS} to reflect their actual deployability into real-world scenarios, we answer four key research questions within this paper.

\textbf{Q1 -- How many attack samples do supervised \acp{IIDS} need?} 

\noindent The training of supervised \acp{IIDS} requires samples of benign \emph{and} malicious data samples.
As an example, the most commonly used dataset in that research area~\cite{Lamberts2023SoK}, the Morris Gas dataset~\cite{Morrisetal2015Industrial}, consists of $274.628$ samples, of which $22\%$ are attacks.
For evaluations, authors usually randomly shuffle and split this dataset, leveraging $80\%$ for training and the rest for evaluations~\cite{anton2019anomaly,Perezetal2018Machine}.
With this split, the training data still contains around $48.000$ attack samples.
Yet, obtaining this amount of attack samples from each \ac{ICS} an \ac{IIDS} should be deployed is unrealistic considering the costs and risks associated with their collection.
We find that supervised \acp{IIDS} are unsuitable for real deployments due to only performing well with many attack samples, potentially due to overfitting, which aligns with prior research~\cite{Ahmedetal2020Challenges,Etalle2017From,Kus2022A,M.-R.etal2021Machine}.

\textbf{Q2 -- How much training data do OCC-based IIDSs need?}
\acp{IIDS} requiring only benign training data can be trained with less difficulty, \eg even during the regular operation of an \ac{ICS}.
However, this training data must still be collected, and it must be ensured that it reflects \emph{all} possible genuine behavior.
Hence, we want to understand how much training data is actually necessary and whether large variances exist across detection methods.

\textbf{Q3 -- What is the influence of hyperparameters on performance?}
Beyond training data, \ac{OCC}-based \acp{IIDS} request hyperparameters, which may significantly impact detection performance.
Here, the MinMax \ac{IIDS} introduced in the background~(\cf \secref{sec:bg:iids}) uses a fixed threshold across datasets, whereas an optimized threshold could drastically influence detection performance, as evidenced in \figref{fig:minmax-example}.
However, such hyperparameter tuning is only possible if attack samples for the concrete deployment scenario are available.

\textbf{Q4 -- Can we transfer good hyperparameters across scenarios?}
To unlock the benefits of tuned hyperparameters in OCC-based \acp{IIDS}, we consider the previously proposed concept of transferring good configurations across deployment scenarios~\cite{probst2019tunability,weerts2020importance}.
Such a step would also allow us to use the extensively available attack samples from artificial scenarios to tune real-world deployments.
However, thus far, it remains unclear to what extent such transferability is possible and to what extent this is scenario and \ac{IIDS} dependent.

\section{Deployability of Supervised IIDS}
\label{sec:training}

Our initial analysis concerns the deployability of supervised \acp{IIDS} \wrt the amount of required attack samples.
We first describe our experiment design, then analyze our results, and finally summarize the implications of our findings.

\subsection{Experiment Setup}
\label{sec:training:experiment}

In the following, we present the \acp{IIDS}, datasets, and conducted experiment methodology to tackle the research question Q1.

\subsubsection{IIDSs.}
\label{sec:training:experiment:iids}
For our experiments on supervised \acp{IIDS}, we consider a \ac{RF} and a \ac{SVM} classifier as used in several proposed \acp{IIDS}~\cite{anton2019anomaly,junejo2016behaviour,Perezetal2018Machine}.
As independently examined by Perez \etal~\cite{Perezetal2018Machine} and Anton \etal~\cite{anton2019anomaly}, these classifiers can be adapted to operate on Modbus network traffic via derived features such as the function code or transmitted process values.
The classifiers are trained and evaluated on a set of benign and malicious Modbus packets.
Our experiment is based on existing re-implementations of these two \acp{IIDS} made available in the IPAL IDS Framework~\cite{Wolsing2022IPAL}.
We took care to use the same data preprocessing and hyperparameters as mentioned in the publication~\cite{Perezetal2018Machine} (\cf Availability Statement).

\subsubsection{Dataset.}
\label{sec:training:experiment:dataset}
We leverage the same dataset originally used to evaluate the two analyzed \acp{IIDS}~\cite{anton2019anomaly,Perezetal2018Machine}, which is also the most commonly used dataset for supervised \acp{IIDS}~\cite{Lamberts2023SoK}.
This dataset has been recorded in a miniature gas-pipeline \ac{ICS} environment leveraging Modbus as communication protocol.
Within this setup, a total of $60048$ attack samples across $35$ types of attacks with varying complexity, such as reconnaissance or modifying setpoints, have been collected.

\subsubsection{Conduction.}
\label{sec:training:experiment:conduction}
To understand how many attack samples are necessary to adequately train a supervised \ac{IIDS},
we reduce the number of samples contained in the training dataset while keeping the number of benign training data constant.
We start with a random 80/20 train/test split and five folds as used for the original evaluation~\cite{anton2019anomaly,Perezetal2018Machine}.
We then remove all but one attack sample from the training data and train new classifiers while gradually increasing the amount of attack samples in the training data.
For each number of learned attack samples, we calculate the average recall (fraction of identified attacks) and precision (fraction of correct alerts) over all folds.

\begin{figure}[t]
	\centering
	\subfigure[After an initialization phase, the recall increases linearly with more attack samples, while changes in precision are only minimal. To yield high detection scores, more than $40.000$ malicious packets are required in training for both supervised \acp{IIDS}. Note that the data for \ac{SVM} was sampled in steps of 500 attacks due to long training times.]{\includegraphics[width=\columnwidth]{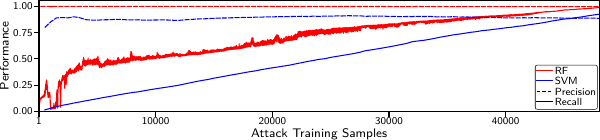}\label{fig:training:all}} \\

	\subfigure[The \ac{RF} requires few training samples on attack number $19$ (clean registers).]{\includegraphics[width=.49\columnwidth]{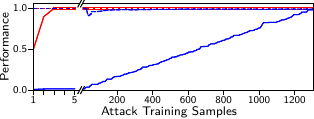}\label{fig:training:easy}}
	\hfill
	\subfigure[Attack $10$ (change physical value) requires many samples to be trained.]{\includegraphics[width=.49\columnwidth]{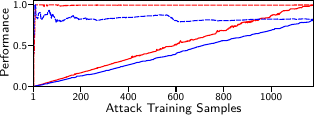}\label{fig:training:hard}}

	\caption{Gradually increasing the amount of attack samples within the training data reveals that both \ac{RF} and \ac{SVM} require lots of data to yield satisfying detection performance. For a simple attack, \cf \figref{fig:training:easy}, the \ac{RF} requires only about three samples.}
	\label{fig:training}
\end{figure}

\subsection{Q1 -- How many attack samples do supervised IIDSs need?}
\label{sec:training:q1}

Having established our evaluation methodology, we can now exemplarily assess the deployability of supervised \acp{IIDS} \wrt to the amount of attack samples.
To this end, \figref{fig:training} depicts the overall detection performance of the \ac{RF} and \ac{SVM} classifiers.
In \figref{fig:training:easy} and \figref{fig:training:hard}, we then see the detection performance reduced to two exemplary attack types.

Starting with a broad overview in \figref{fig:training:all}, if the \ac{RF} is trained on all available attack samples ($x = 48.049$), it reaches a precision of $\evalSupervisedRFFinalPrecision$ and a recall of $\evalSupervisedRFFinalRecall$.
Likewise, the \ac{SVM} achieves a score of $\evalSupervisedSVMFinalPrecision$ in precision and $\evalSupervisedSVMFinalRecall$.
As we expect, both \acp{IIDS} achieve the best detection performance when trained on all (those in the original train set) available attack samples.

However, as we reduce the number of attack samples, we observe a nearly linear reduction in recall of both \acp{IIDS}.
For \ac{RF}, the performance drops from $\evalSupervisedRFFinalRecall$ to about $0.44$ if provided with just $5000$ attack samples.
Below this threshold, the recall for \ac{RF} drops even more drastically.
The \ac{SVM} shows a similar trend while showing fewer fluctuations.
Interestingly, precision remains largely unaffected in both cases, which we presume results from not changing the amount of benign training data.

To better understand these effects, we now conduct the same experiment while only considering a single attack type during training and testing.
Exemplarily, we show attack type $19$ (as defined for the dataset~\cite{Morrisetal2015Industrial}) in \figref{fig:training:easy} and attack type $10$ in \figref{fig:training:hard}.
For attack type $19$, we observe a vastly diverging behavior between \ac{RF} and \ac{SVM}.
The \ac{RF} achieves optimal detection rates after just three attack samples.
Here, the \ac{IIDS} has likely learned to identify that this attack uses a Modbus function code not occurring during normal behavior.
In contrast, this generalization does not apply to the \ac{SVM}.
Attack type $10$, shown in \figref{fig:training:hard}, which manipulates reported sensor readings, proves difficult to learn for both \acp{IIDS}.
Here, the recall continues to grow linearly as more attack samples are available for training.
Overall, we see that only with a high number of malicious training samples can the \acp{IIDS} score the excellent detection results reported in the respective publications.

\subsection{Conclusion}
\label{sec:training:conclusion}

Looking back at our results, we see that supervised \acp{IIDS} can generalize an attack pattern in some cases as observed, for example, for the \ac{RF} classifier for attack type $19$, which introduces an otherwise unused Modbus function code.
This attack should thus also easily be detected by simple rule-based \acp{IIDS}~\cite{Fovinoetal2010Modbus/DNP3}.
In general, however, we observe relatively little generalization for both \acp{IIDS}.
The linearly increasing recall scores with increasing the number of attack samples rather indicate an overfitting behavior of the classifiers, \ie only the precise misbehavior observed during training is also later classified as such.
These results provide further evidence for prior work by Kus \etal~\cite{Kus2022A}, who already identified a lack of generalization during supervised \ac{IIDS} training.

All in all, the prospects for supervised \acp{IIDS} look rather grim.
They require an unrealistic high number of attack samples for training and then do not even generalize malicious behavior.
Consequently, novel designs for supervised \acp{IIDS} must be researched to be realistically considered for real \ac{ICS} deployments.
Shockingly, these issues are hardly discussed in prior work as publications promoting the use of machine-learning in \ac{ICS} mainly focus on the final achieved detection performance~\cite{anton2019anomaly,junejo2016behaviour,Perezetal2018Machine}.

\section{Deployability of OCC-based IIDS}
\label{sec:stability}

\ac{OCC}-based \acp{IIDS} promise to circumvent these issues of supervised \acp{IIDS} by requiring only training data from benign \ac{ICS} operations.
Getting such benign data is easier than collecting attack samples, but it must still be collected, processed, and verified, such that requiring less training data makes an \ac{OCC}-based \ac{IIDS} easier to deploy.
Moreover, hyperparameter tuning, especially if hyperparameters cannot be transferred across scenarios, can still unrealistically boost an \ac{OCC}-based \acp{IIDS}' performance in research.
To understand these effects, we first lay out the evaluation setup underlying our measurements  (\secref{sec:methodology}) to then tackle the research questions Q2 to Q4.
In the end, we summarize our findings on the deployability challenges of \ac{OCC}-based \acp{IIDS} (\secref{sec:stability:conclusion}).

\subsection{Experiment Setup}
\label{sec:methodology}

To holistically conduct our experiments, we select four \acp{IIDS} from related work and three frequently used datasets.
We then define the examined hyperparameter space for which we ultimately measure their performance.

\subsubsection{IIDSs.}
\label{sec:methodology:iidss}
We examine four \acp{IIDS} designed for industrial use cases, which were published at top security conferences.
When evaluating these \acp{IIDS}, we make use of available open-source implementations or validated re-implementations within the IPAL \ac{IIDS} framework~\cite{Wolsing2022IPAL}.
In the following, we briefly lay out the detection idea of each approach, but for further details, we refer the reader to the respective publications or IPAL's public implementation~\cite{Wolsing2022IPAL}.

\vspace*{2mm}
\textit{MinMax.}
The first \ac{IIDS}, MinMax, learning the minimum and maximum bounds of a sensors' normal values (\cf \secref{sec:bg:iids}), serves as a representative for a class of lightweight \acp{IIDS} that aim to implement straightforward detection methodologies that do not require complex configuration, technical understanding, or computational resources~\cite{Wolsing2022Can}.
Any violation against the learned minimum and maximum values is indicated as an alert to the \ac{ICS} operators.

\vspace*{2mm}
\textit{Invariant.}
The Invariant \ac{IIDS}~\cite{Feng2019A} leverages data mining techniques to find mathematical equations that must be fulfilled at all times.
E.g., if the inlet valve of a water tank is opened, its water level is expected to rise.
Since these invariants are fulfilled all the time during normal behavior, any violation of such a rule is then reported as an alert.

\vspace*{2mm}
\textit{TABOR.}
This \ac{IIDS} fuses three detection approaches based on timed automata, Bayesian networks, and out-of-bounds checks~\cite{Lin2018TABOR}.
The timed automata component considers a single sensor value and learns a model of its behavior.
E.g., the water levels of a tank usually rise for 30 minutes and then decrease over several hours.
Together with the Bayesian network, unknown process states can be determined, such as the inlet valve being still opened despite the water level rising for more than 30 minutes.
To complement their method, an alert is also raised with an out-of-bounds check working similarly to the MinMax \ac{IIDS}.

\vspace*{2mm}
\textit{Seq2SeqNN.}
Lastly, Seq2SeqNN~\cite{Kim2020Anomaly} trains a neuronal network on GPUs to understand the \ac{ICS}'s behavior and perform predictions for the future.
Given a recent history of physical values, the neuronal network is able to perform a prediction for the near future.
If these predictions deviate too much from the observed behavior, an alarm is raised.

\vspace*{2mm}
These four \acp{IIDS}, which also feature vastly different numbers of hyperparameters for their configuration (\cf \tabref{tab:setup}), build the foundation for our  analysis.

\begin{table}[t]
	\caption{We analyze four state-of-the-art \acp{IIDS} with diverse hyperparameters, on three datasets. We aim at 10.000 random samples for each \ac{IIDS}'s hyperparameter space, yet we have reached computational limits, resulting in fewer samples for some.}
	\label{tab:setup}

	\centering
	\begin{tabular}{ l rrrr}
		\textbf{\ac{IIDS}} & \multicolumn{1}{c}{\textbf{SWaT~\cite{Gohetal2016A}}} & \multicolumn{1}{c}{\textbf{WADI~\cite{Ahmedetal2017WADI}}} & \multicolumn{1}{c}{\textbf{BATADAL~\cite{Taorminaetal2018Battle}}} & \multicolumn{1}{c}{\textbf{Parameter}} \\
		\toprule

		MinMax~\cite{Wolsing2022Can} & \num{\evalMeasurementsSWaTSIMPLEMinMax} & \num{\evalMeasurementsWADISIMPLEMinMax} & \num{\evalMeasurementsBATADALSIMPLEMinMax} & \num{\evalParametersSIMPLEMinMax} \\
		Invariant~\cite{Feng2019A} & \num{\evalMeasurementsSWaTInvariant} & \num{\evalMeasurementsWADIInvariant} & \num{\evalMeasurementsBATADALInvariant} & \num{\evalParametersInvariant} \\
		TABOR~\cite{Lin2018TABOR} & \num{\evalMeasurementsSWaTTABOR} & \num{\evalMeasurementsWADITABOR} & \num{\evalMeasurementsBATADALTABOR} & \num{\evalParametersTABOR} \\
		Seq2SeqNN~\cite{Kim2020Anomaly} & \num{\evalMeasurementsSWaTSeqSeqNN} & \num{\evalMeasurementsWADISeqSeqNN} & \num{\evalMeasurementsBATADALSeqSeqNN} & \num{\evalParametersSeqSeqNN} \\
		\bottomrule
	\end{tabular}
\end{table}

\subsubsection{Datasets.}
\label{sec:methodology:datasets}

To generalize our results, we analyze each \ac{IIDS} on three popular datasets, namely the SWaT~\cite{Gohetal2016A},  WADI~\cite{Ahmedetal2017WADI}, and BATADAL~\cite{Taorminaetal2018Battle}, which are among the most commonly used datasets in this research area~\cite{Lamberts2023SoK}.
All three datasets come with dedicated training data that is free of attacks.
SWaT and WADI have one evaluation dataset containing 36 and 14 different cyberattacks, respectively.
BATADAL has two evaluation datasets with five and seven attacks, respectively, for which discuss the concatenated results.

\subsubsection{Hyperparameter Selection.}
\label{sec:methodology:parameters}

While the previous experiment's design decisions coincide with usual \ac{IIDS} evaluation methodologies~\cite{Lamberts2023SoK}, our work differs within the hyperparameter selection we aim to study.
Although three of the examined \acp{IIDS}' publications contain short discussions about (some) hyperparameters~\cite{Feng2019A,Kim2020Anomaly,Wolsing2022Can}, none defines the precise acceptable range of the hyperparameter space.
To this end, we have to come up with our own definition.
For nominal and ordinal hyperparameters, we simply enumerated all possible values, and for rational numbers, we had to define a custom range based on our understanding of the proposed system.
During their definition, we took special care that the values proposed in the original publications are contained in our analyzed ranges.

\subsubsection{Conduction.}
\label{sec:methodology:setup}

Finally, to conduct a parallelized examination of the hyperparameter in a repeatable manner, we leveraged Ray Tune~\cite{Liawetal2018Tune}, a library to scale hyperparameter search and tuning.
Provided with a definition of a hyperparameter search space, Ray Tune selects one hyperparameter configuration uniformly at random at a time and then trains and evaluates the respective \ac{IIDS} on the dataset.
We then calculate the precision, recall, and F1 score metrics, as these are among the most common performance metrics in \ac{IIDS} research~\cite{Lamberts2023SoK}.

As shown in \tabref{tab:setup}, we achieved up to $10.000$ samples for the different \acp{IIDS} and datasets, building a solid foundation for our subsequent analyses.
In some cases, such as evaluating the Invariant \ac{IIDS} on SWaT, training a single configuration takes up to eleven days, which explains the reduced number of samples.
Similarly, the training of the Seq2SeqNN \ac{IIDS} requires exclusive access to potent GPUs to train a neural network.
To grant other researchers access to the result of these extensive computations for further analyses, we made all collected data publicly available, \cf Availability Statement.

\subsection{Q2 -- How much training data do OCC-based IIDSs need?}
\label{sec:stability:q2}

First, we want to understand the impact of the amount of (benign) training data on the \acp{IIDS}' performance. 
Here, we only consider the best hyperparameters found \wrt the F1 score for each \ac{IIDS} and dataset combination.
Beginning with the entire training data (100\%), we gradually reduced the training data and evaluated the \ac{IIDS} after each training against the entire test dataset.

\begin{figure}[t]
	\centering
	\includegraphics[width=\columnwidth]{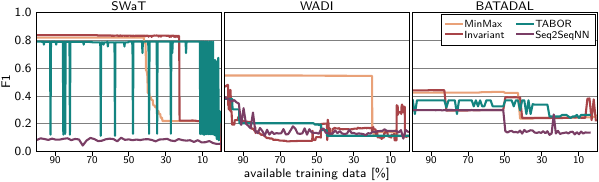}
	\caption{Reducing the amount of benign training data likewise diminishes the detection rate but not all \acp{IIDS} experience an equal performance reduction. The training data Seq2SeqNN on SWaT and WADI is reduced in steps of $10.000$ in contrast to $1.000$ for the others whereas all \acp{IIDS} on BATADAL are sampled in steps of $100$.}
	\label{fig:decreasing}
\end{figure}

As shown in \figref{fig:decreasing}, the amount of training data impacts the detection performance of each \ac{IIDS} differently.
E.g., the performance of MinMax on SWaT and BATADAL initially stays high.
Only when the data is reduced to about less than $40\%$ does the performance drop significantly.
On WADI, this drop occurs much later at about $20\%$ of the overall training data.
For Invariant, we observe a similar pattern on SWaT achieving top scores even with about $25\%$ of the data.
Yet, on WADI, this approach requires nearly all training data to get close to its optimal score.
TABOR on SWaT shows another interesting behavior where instead of a slow reduction, we observe occasional drops in performance, which accumulate toward the end.
Upon investigation of TABOR's trained model, we noticed that the drops in between are caused by learning a different model, showing the unstable nature of the trained model.
This also occurs in reduced form for BATADAL but not on WADI, where TABOR shows a more continuous reduction as less training data is made available.
Seq2SeqNN performs poorly on SWAT and on the other datasets its performance drops significantly as training data is reduced to $50\%$.
Overall, we observe that all \acp{IIDS} perform nearly optimally on SWaT and BATADAL with just about half of the training data, while performance on WADI often quickly drops off.

\textbf{Takeaway.}
Our data shows that judging upfront whether one has acquired enough training data in a deployment scenario can be challenging.
The amount of necessary training data seems to be neither directly dependent on the \ac{IIDS} nor on the complexity of the concrete scenario.
As Invariant and Seq2SeqNN on WADI experience a substantial increase close to $100\%$ training data, this may be an indication that these IIDS would benefit from even more training data than contained in the dataset.
We also see that the performance of the different \acp{IIDS} drops suddenly after a certain point, indicating that not observing some specific event during training can be responsible for much of the performance loss.
Interestingly, the different \acp{IIDS} seem to have different events triggering their performance loss.
When interpreting these results, dataset characteristics should also be kept in mind.
E.g., SWaT contains one attack that is significantly longer than the others, which significantly worsens the F1-score if it is not detected anymore.
Hence, the sudden drops of MinMax and Invariant on SWaT could be explained by the sudden inability to identify that specific attack.
Overall, we can, however, say that determining the amount of necessary training data varies across \acp{IIDS} and scenarios, such that a final assessment can only be made on a case-by-case basis.

\subsection{Q3 -- What is the influence of hyperparameters on performance?}
\label{sec:stability:q3}

\begin{figure}[t]
	\centering
	\includegraphics[width=\columnwidth]{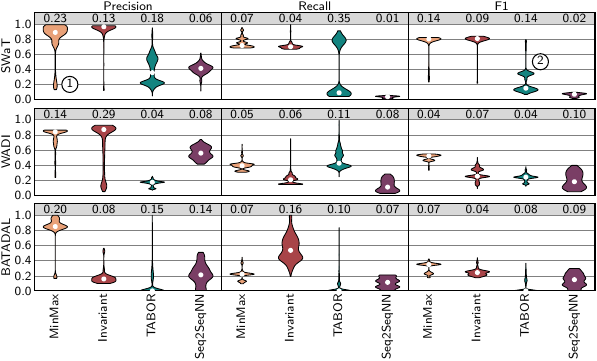}
	\footnotesize{The white dot represents the median and the numbers on top are the standard deviation of each distribution.}

	\caption{
	An \acp{IIDS}' performance depends on an optimal choice of hyperparameter.
	While MinMax or Invariant yield satisfying results in F1 score on SWaT for the majority of configurations, obtaining a good configuration for TABOR is challenging.
	Thus, judging the expectable performance of an \ac{IIDS} by a single number can be misleading.
	}
	\label{fig:boxplots}
\end{figure}

Next, for our investigation on the significance of hyperparameters (Q3), we take a broad view of the obtained measurements (\cf \secref{sec:methodology}).
To this end, \figref{fig:boxplots} depicts every \ac{IIDS}'s performance distribution along several metrics and datasets.

At first glance, we observe that hyperparameters have a tremendous effect on the performance of \acp{IIDS}.
E.g., considering the precision of the MinMax \ac{IIDS} on the SWaT dataset~(\cf \circled{1} in \figref{fig:boxplots}), the performance varies between \num{\evalMaxSWaTPrecisionSIMPLEMinMax} at best and \num{\evalMinSWaTPrecisionSIMPLEMinMax} at worst, which implies that, depending on the chosen configuration, the approach performs close to optimal or is inapplicable.
But looking at the entire distribution, it becomes apparent that low values in recall are outliers as the median performance~(white dots) is still high at \num{\evalMedianSWaTPrecisionSIMPLEMinMax}.
Still, the standard deviations around the median is relatively high at \num{\evalStdSWaTPrecisionSIMPLEMinMax}, and thus, performance penalties can be expected for MinMax in recall if not parameterized correctly.

Taking a broader look at the precise distribution of different approaches, not all \acp{IIDS} exhibit the same patterns.
When considering MinMax and Invariant for SWaT in the F1 score, the majority of configurations perform decently, and bad results are mostly outliers.
We call this type of distribution \emph{stable} as it is quite likely to pick a good-performing configuration without having to invest great efforts.
In contrast, the opposite is true for TABOR \circled{2}, with a median of just \num{\evalMedianSWaTFTABOR}, which is far from what could be achieved at best (\num{\evalMaxSWaTFTABOR}) with this approach.
Here, unlike MinMax, it is quite unlikely to hit such a good-performing configuration even with expert knowledge.
Therefore, there is a qualitative difference between the presumably stable MinMax, which promises to have a straightforward configuration process~\cite{Wolsing2022Can}, and TABOR.
Note that for MinMax, these observations may be affected by only having two hyperparameters in the first place~(\cf \tabref{tab:setup}).
Still, Invariant, despite having the most parameters, features a similar stable distribution as MinMax, at least \wrt the F1 score.

Next, we want to understand whether the (in-)stability property is inherent to a specific \ac{IIDS}.
First and foremost, note that the absolute scores achieved between the datasets (\cf lower part of \figref{fig:boxplots}) are sometimes lower compared to SWaT as not all \acp{IIDS} were primarily designed for the other datasets.
Hence, we only focus on the distributions here.
In general, the distributions are loosely similar in each setting.
The performance distributions of most \acp{IIDS} have roughly the same features on WADI and BATADAL, with some exceptions, such as WADI missing the outliers to the top in some cases.
This observation indicates that the stability of an \ac{IIDS} may be dominantly determined by the underlying detection mechanism rather than the scenario.
Consequently, stability seems to be an inherent feature of an IIDS, which could act as a proxy for determining how easy or difficult deploying an \ac{IIDS} in a new, real application may be.

Considering MinMax, the authors publish their \ac{IIDS} with a F1 score of \num{0.78} for SWaT and \num{0.52} for WADI~\cite{Wolsing2022Can}.
W.r.t. our evaluation, these numbers are close to the median (SWaT \num{\evalMedianSWaTFSIMPLEMinMax} and WADI \num{\evalMedianWADIFSIMPLEMinMax}) and leave headroom to the maximum (\num{\evalMaxSWaTFSIMPLEMinMax} respectively \num{\evalMaxWADIFSIMPLEMinMax}).
Thus, the published numbers are representative of the expectable performance, which comes as no surprise as the authors stated not to have performed any parameter optimization~\cite{Wolsing2022Can}.
In contrast, Feng \etal~\cite{Feng2019A} promote the Invariant \ac{IIDS} with a recall of \num{0.79} for SWaT and \num{0.47} for WADI.
Compared to the median performance (SWaT \num{\evalMedianSWaTRecallInvariant} and WADI \num{\evalMedianWADIRecallInvariant}), the published values are outliers by multiple standard deviation.
Therefore, it can be assumed that Feng~\etal published optimized performance statistics.
Such fine-tuning certainly has value in examining what maximal performance can be achieved by a proposed approach.
However, such results carry the risk of misrepresenting how good a system may perform in a real deployment and may prevent fair comparisons of approaches.

\begin{figure}[t]
	\centering
	\includegraphics[width=\columnwidth]{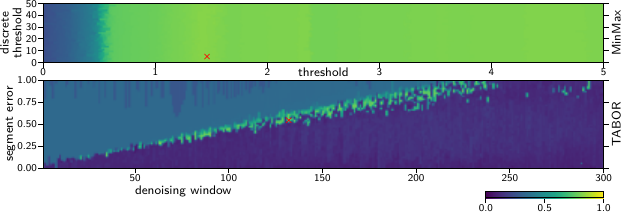}
	\footnotesize{The red cross (\red{$\times$}) indicates the optimum found during the experiment.}

	\caption{The impact of hyperparameters can vary significantly between approaches. While on the SWaT for MinMax (upper plot), one parameter is decisive for the entire performance, suitable configurations for TABOR (lower plot) are more challenging to obtain as several parameters influence each other.}
	\label{fig:deepdive}
\end{figure}

Given these distinct behaviors, \acp{IIDS} show under varying hyperparameters, we also have to ask what the reasons for these behaviors might be.
Therefore, we take a closer look at MinMax and TABOR on the SWaT dataset and F1 score, as depicted in \figref{fig:deepdive}, where we visualize the detection performance as a heatmap in dependence of two relevant hyperparameters.
For MinMax, we identify that the final result mostly depends only on the threshold parameter~(\cf \secref{sec:bg:iids}).
For small values (below \num{0.6}), there is a significant drop in detection performance, but afterward, there are only subtle changes and the threshold has no significant impact anymore.
In contrast, for TABOR, we observe more interdependence in two of the seven hyperparameters.
Both parameters influence the performance, and changes to one parameter alter the optimal value of the other parameter.
Thus, only a combination of correctly set hyperparameters yields good configurations, which complicates setting up TABOR, which explains our previous observation where only a few configurations yielded good performance.

\textbf{Takeaway.}
We observed that hyperparameters have a tremendous impact on the measured performance of \ac{OCC}-based \acp{IIDS}.
Moreover, there exist considerable differences in \ac{IIDS} stability.
The MinMax or Invariant \acp{IIDS} yield results that are close to their optimal in a majority of configurations.
At the same time, TABOR only achieves optimal performance if multiple hyperparameters are fine-tuned.
Our results stand in contrast to Fung \etal, who claimed that reconstruction-based \acp{IIDS} can have a good performance over a broad spectrum of hyperparameters~\cite{Fungetal2022Perspectives}, likely because our evaluation covered a more diverse set of \acp{IIDS}.
This (in-)stability \wrt hyperparameters complicates scientific comparisons and real-world applicability if the performance of an \ac{IIDS} is only acceptable for a very confined parameter space.
Consequently, we warn that judging an \acp{IIDS}' performance by a single configuration, as done currently throughout the literature, can be misleading.

\subsection{Q4 -- Can we transfer good hyperparameters across scenarios?}
\label{sec:stability:q4}

As we discussed in the previous section, it can be difficult to obtain suitable hyperparameters for an \ac{IIDS} for a given deployment or dataset.
For the selection of suitable hyperparameters, we do, however, not need to start from scratch in most cases.
Instead, published parameters or guidelines from previous deployments may help to identify good parameters.
Thus, one idea is to reuse these already known configurations and transfer them to a new scenario to hopefully achieve adequate performance.
If such hyperparameter transfers are feasible, it would alleviate the problem of (in-)stability discussed before.
Previously, Probst~\etal~\cite{probst2019tunability} found universally good-performing default hyperparameters for supervised \acp{IIDS}.
However, we consider \acp{OCC}-based \acp{IIDS} with potentially more intricate hyperparameters that may hinder such transferability.

\begin{figure}[t]
	\centering
	\includegraphics[width=\columnwidth]{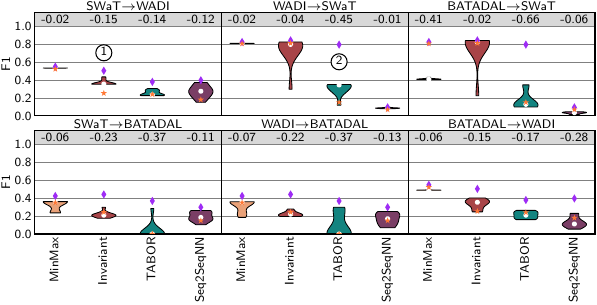}
	\footnotesize{The numbers on top are the difference between the median (white dot) and maximum achievable performance (\markdiamond).}

	\caption{
	Transferring the top ten configurations found for one dataset to another promises to avoid the problem of parameter optimization in new settings.
	But, this methodology usually lacks behind the achievable optimum (\markdiamond) and does not systematically exceed randomly selected hyperparameters, \cf\xspace \markstar marking the median of all hyperparameters in the target dataset.
	}
	\label{fig:transfer}
\end{figure}

As the first step in that direction and to examine whether a known, good-performing configuration is also suitable on a different dataset, we conducted the following evaluation.
First, we select the top ten configurations according to the F1 score of an \ac{IIDS} and dataset, \eg MinMax on SWaT, and measure the performance of these hyperparameters applied to the other datasets, \ie WADI and BATADAL.
Applying this methodology, \figref{fig:transfer} depicts the distribution of the obtained ten results on the new datasets.
In addition, we mark the globally achievable optimum in a given setting found in the previous analysis from \secref{sec:stability:q3} with an \markdiamond and the median performance of randomly selected hyperparameters with an \markstar.
This analysis enables assessing how likely a configuration is transferable to another scenario without tuning but also the potential losses in performance along the way.
Note that since we sampled all hyperparameters randomly (\cf \secref{sec:methodology}), it is not guaranteed that we measured the precise configuration on the respective other datasets.
In that case, we selected the measurement closets to the selected default configuration.

We start considering the Invariant \ac{IIDS} transferred from the SWaT to the WADI dataset as a case study first (\cf \circled{1} in \figref{fig:transfer}).
We see that the transferred configurations achieve decent performance relative to the achievable maximum of \num{\evalMaxWADIFInvariant}.
The median expectable performance from transferring the configurations (white dot) is \num{\evalTransferDiffSWaTWADIInvariant} points lower than the maximum achievable.
Given that no effort was required to find these configurations, this median of transferred configurations \num{\evalTransferMedianSWaTWADIInvariant} is an improvement over the previous random median performance (\num{\evalMedianWADIFInvariant}, \cf \figref{fig:boxplots}).
In contrast to this example, there also exist cases where hardly any transferability is possible.
When considering TABOR transferred from the WADI to the SWaT dataset (\cf \circled{2} in \figref{fig:transfer}), there is a large gap between the median transferred configuration (\num{\evalTransferMedianSWaTWADITABOR}) and the achievable optimum (\num{\evalMaxSWaTFTABOR}).
While in that case, transferring the results is still better than drawing a random configuration (\num{\evalMedianSWaTFTABOR}), a large potential is left on the table.
More generally, the median transferred performance (white dots) is, on average, \num{\evalAvgDistUnsup} lower than the respective achievable optimum (\markdiamond).
At the same time, this method is equal to randomly selecting a configuration (no difference to \markstar on average).
Thus, while transferring configurations from one scenario to another seems promising, this concept still proves not be be that advantageous.

\textbf{Takeaway.}
On the one hand, \ac{OCC}-based \acp{IIDS} lack guidelines for hyperparameter configuration.
On the other hand, if known configurations exist, they  only offer limited transferability to new scenarios.
On average, good hyperparameters on one dataset do not perform better than randomly chosen hyperparameters on another dataset.
Thus, if an \ac{IIDS} is challenging to configure in the first place, even default configurations or templates from other scenarios do not help much, and, in the worst case, manual efforts are required to tune the approach individually.

\subsection{Conclusion}
\label{sec:stability:conclusion}

We began with the observation that the amount of data required for training differs significantly between \acp{IIDS} (Q2), which complicates providing concrete advice for deployments.
Still, as one redeeming feature, more training data does not seem to negatively impact the detection performance.
Next, we studied the hypothesis that hyperparameters are a crucial factor for \ac{IIDS} performance and again observed vastly different behaviors \wrt stability.
Indeed, it proved difficult for some \acp{IIDS} to yield good results on average (Q3), and quick solutions such as generic default configurations that generalize to new scenarios or datasets did not prove promising (Q4).
In contrast to the works by Probst \etal~\cite{probst2019tunability} and Weerts \etal~\cite{weerts2020importance}, we found that deriving default values for hyperparameters of \ac{OCC}-based \acp{IIDS} for the \ac{ICS} domain is challenging for our three analyzed datasets and tuning them manually based on attack samples still brings an enormous performance benefit.
Therefore, these effects can explain previously reported problems from related work, \eg failed reproducibility studies or deploying such approaches in practice~\cite{Ahmedetal2020Challenges,Etalle2017From,Sommeretal2010Outside}.
I.e., Erba \etal~\cite{Erbaetal2020No} tried to reproduce the Invariant \ac{IIDS} and had troubles finding the hyperparameters to mach the publications result.
This is in line with our assumption from \secref{sec:stability:q3} where we presumed that the authors of the Invariant \ac{IIDS} have tuned their published parameters.
Consequently, current evaluation methodologies in research omit a relevant attribute of \acp{IIDS} that is currently not easily measurable.

In general, obtaining a quantitative intuition on an \ac{IIDS}'s training and tuning demands can provide valuable data on the one hand, for \ac{ICS} operators having to select, set up, and configure an \ac{IIDS} and, on the other hand, for research to establish fairer and easier comparisons.
Note that we do not want to prioritize an \ac{IIDS} with low training and low tuning demands over ones with excellent detection performance.
Instead, we want to create awareness for these challenges and advocate for researchers to scrutinize their work more \wrt their deployability.

\section{Open Deployment Issues and Call to Action}
\label{sec:call}

Our results regarding the analysis of research questions Q1 to Q4 prominently show that there exist complex challenges to transferring an \ac{IIDS} developed in research to an actual \ac{ICS} that are not captured accurately by the current standard in \ac{IIDS} evaluations.
Hence, the standard procedure of publishing the detection performance for one or multiple datasets~\cite{Lamberts2023SoK} is insufficient to capture an \ac{IIDS}' true value.
Concerning these issues, in this section, we discuss  potential new strategies to assess the ease and limitations of an \ac{IIDS}' deployment already during the research stage.

One significant obstruction in deploying \acp{IIDS} is acquiring sufficient training data (Q1 and Q2) whilst avoiding overfitting of supervised \ac{IIDS} models.
From a research perspective, an adopted evaluation methodology that more deeply assesses the capabilities and especially training properties of an \ac{IIDS} in the lab may be suited to estimate its training demand upfront.
In that regard, the evaluation methodologies we presented enable, one the one hand, inferring the learning rate from which the amount of required training data for a deployment can be estimated.
On the other hand, by visualizing the learning rate of individual attacks, first signs of overfitting can be revealed.
In addition, the methodology proposed by Kus \etal~\cite{Kus2022A} can answer how well a supervised approach generalizes to unknown attacks, \eg found during live operations, which are not part of the training data.
Together, such enhanced evaluation methodologies can reveal \acp{IIDS} that a) require little training samples and b) generalize to a wide variety of (zero-day) cyberattacks beyond the ones seen in training.

With the previous issues addressed, the challenge of configuring an approach (Q3 and Q4) remains.
For research, analyzing (new) \acp{IIDS} \wrt their stability in hyperparameters or ease of configuration, as done by us, can provide additional information for \ac{ICS} operators on which \ac{IIDS} approach may be best suited for a given deployment.
Therefore, we ideally need a compact metric that expresses the average performance or stability of performance results.
While the data generated in our publication would allow us to compute such values (\cf \figref{fig:boxplots}), how to arrive at a holistic metric that is adequate for scientific purposes is still unclear to us.
Another idea for better understanding \ac{OCC}-based \acp{IIDS} is to use a few attack samples from reference attacks to configure hyperparameters.
Whether this yields good hyperparameters to detect other attacks remains to be seen.

More generally, while the previously sketched concepts for \ac{IIDS} research may work well for research, it is not directly apparent how their insights transfer to actual deployments.
Also, deployability, in general, involves more than recording training datasets and configuring hyperparameters.
E.g., the issue of operational drifts such as wear and tear, which can invalidate once-trained models over time, has been neglected thus far by us~\cite{M.-R.etal2021Machine}.
Answering whether an \ac{IIDS} is ultimately deployable in an actual system thus likely has to involve the expertise of \ac{ICS} stakeholders as already demanded in meta-surveys, \eg by Lamberts \etal~\cite{Lamberts2023SoK}.
Regarding our work, we can, therefore, not finally argue how much training data would still be acceptable or how many false positives and false negatives are tolerable without conducting experiments together with \ac{ICS} experts within an actual \ac{ICS}.

\section{Conclusion}

\acp{ICS} become an indispensable building block for our modern society and, with their high level of digitalization, face potentially disastrous cyberattacks.
As a reaction, research to automatically detect such intrusions took off within the last decade~\cite{Lamberts2023SoK}, and nowadays, with plenty of promising \acp{IIDS}, the transition to deploying those solutions in real-world \acp{ICS} is urgently needed.
Yet, this step involves its own challenges, which we try to identify and quantify in this paper.
Especially the acquisition of adequate training data, avoiding overfitting during training, and the configuration of hyperparameters for \acp{IIDS} to match their excellent detection performance found in (synthetic) research environments is challenging.
As we show, too little training data or tuning of hyperparameters can lead to devastating performance penalties.

While finding solutions to those issues would require the involvement of \ac{ICS} stakeholders that ultimately deploy \acp{IIDS}, we, from a research perspective, recommend taking those properties into account while evaluating novel approaches.
Thereby, we can hopefully shift these deployability challenges more into the focus of researchers who design new intrusion detection methods.

\end{document}